# THE UNIVERSITY OF TEXAS MILLIMETER WAVE OBSERVATORY

## Paul A. Vanden Bout
*National Radio Astronomy Observatory, Charlottesville, VA 22903, USA.*
E-mail: pvandenb@nrao.edu

## John H. Davis
*Department of Electrical Engineering, University of Texas, Austin, TX 78712, USA.*
E-mail: jhd@mail.utexas.edu

and

## Robert B. Loren
*P.O. Box 2915, Silver City, NM 88062, USA.*

**Abstract:** This is an account of the Millimeter Wave Observatory (MWO), a 4.9 m diameter antenna facility that pioneered continuum observations of planets and spectroscopy of interstellar molecules from 1971 tto 1988. The circumstances of its founding, development of its instrumentation, and major research contributions are discussed. The MWO role in training of personnel in this new field is illustrated by a listing of student and postdoctoral observers, with titles of PhD theses that included MWO data.

## 1 INTRODUCTION

The discovery in the late 1960s of interstellar ammonia, water, and hydroxyl, leading to the discovery in 1970 of interstellar carbon monoxide, opened an era of galactic exploration that continues to the present time. The detection of many molecular species, now well over one hundred, revealed a rich and complex chemistry. Molecular spectroscopy proved to be a powerful tool for probing physical conditions in the Galaxy's dark, dense, star-forming interstellar clouds. Although the presence of three molecular radicals in tenuous interstellar gas had been known since the 1930s, spectroscopy of galactic molecular gas did not develop as a major field in astronomy until the exploitation of the radio spectrum, and in particular, the millimeter spectrum. As this development was largely unanticipated, astronomers had to adapt existing telescopes, built for other purposes, to this new area of research. One such telescope was the 4.9-meter diameter millimeter-wave antenna of the Electrical Engineering Research Laboratory (EERL) at the University of Texas (UT) in Austin. For seventeen years, from 1971 to 1988, the Millimeter Wave Observatory (MWO), made pioneering studies of interstellar molecular phenomena. The MWO was able quickly to come on line, a consequence of a number of happy circumstances, not the least being the existence of the antenna itself.

## 2 THE ELECTRICAL ENGINEERING RESEARCH LABORATORY

The Electrical Engineering Research Laboratory (or EERL) was founded in 1942 as an annual research unit of the University of Texas at Austin (UT). Funding for the EERL came primarily from the Department of Defense, which provided generously for over twenty years of research in radio communications, atmospheric propagation, scattering and general electromagnetics. E. Hamlin was the founding Director, but was soon succeeded by Archibald (Archie) Straiton; both were Department of Electrical Engineering faculty members. Straiton's research was guided, in part, by the general trend to higher frequencies in radio communications and applica-tions. By about 1960 he had acquired an interest in astronomy. Perhaps he was inspired by the launch of Sputnik, perhaps by the growing prominence of McDonald Observatory (McD) at the University. It is also possible that he heard about radio astronomy at conferences; many of the first radio astronomers were electrical engineers. Whatever the inspiration may have been, he applied to NASA for support for the construction of a high-quality antenna to study the planets at millimeter wavelengths.

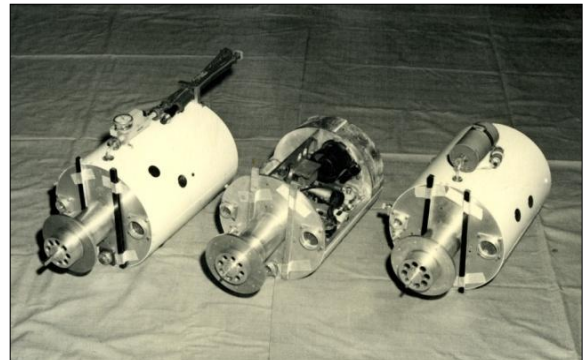

Figure 1: The first receivers of the 4.9-meter antenna, which were used to make planetary brightness measurements from the original site of the antenna, at the Balcones Research Center in Austin (courtesy: files of the EERL).

In 1961 the EERL received two NASA grants, a small ($7,000) grant supporting planetary observations and a larger ($444,000) grant for the antenna. The Western Development Labs, Philco Corp., Palo Alto, California, a subsidiary of the Ford Motor Co., received the contract to build the antenna, and in June of 1963 its construction at the UT Balcones Research Center, on what was then the northern border of Austin and where the EERL was located, was complete. Charles Tolbert, who directed the antenna's construction and its early research program, wrote a description of the new facility (Tolbert, et al. 1965). The first planetary observations at 35, 70, and 94 GHz were of Venus, the brightest of the planets (Tolbert and Straiton, 1964). The receivers, used at prime focus, are shown in Figure 1. Observations of

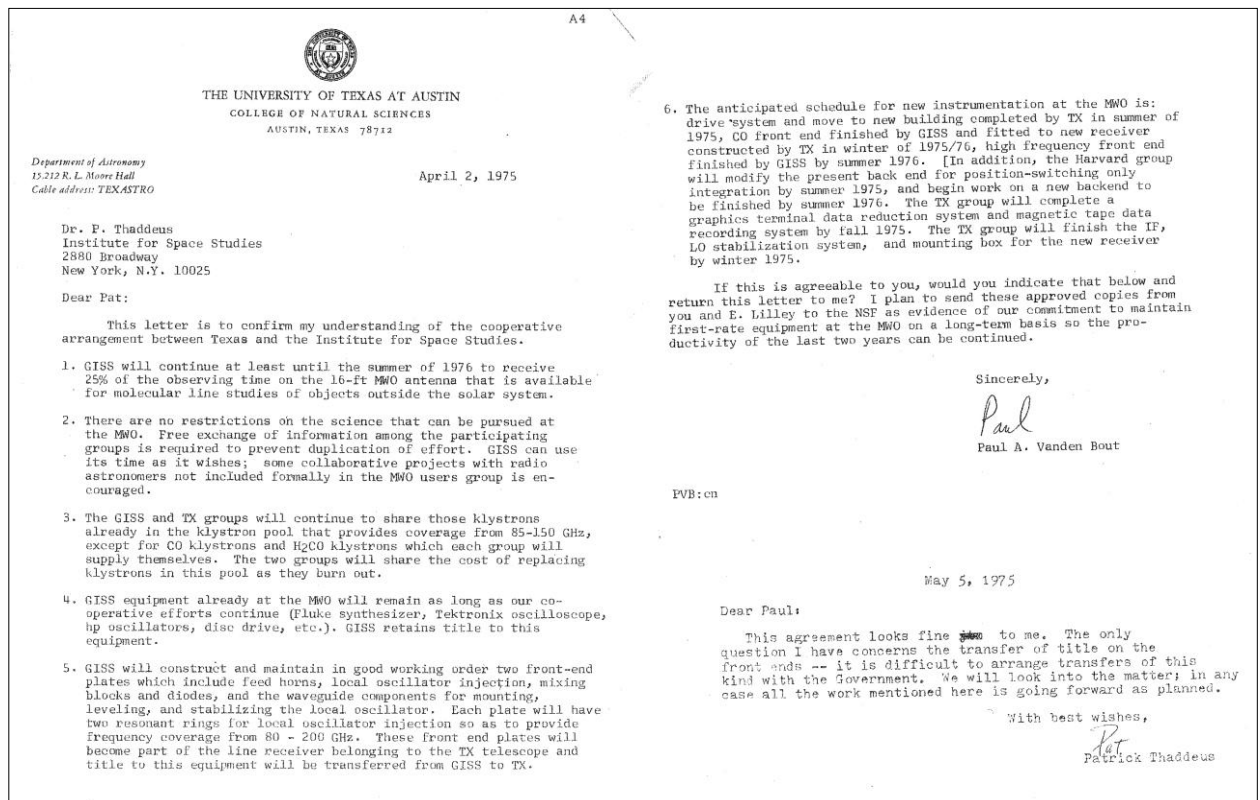

Figure 2: Copy of the letter of agreement for the MWO partnership, signed by P. Thaddeus (from the personal papers of R. Loren).

Mars, Jupiter, and Saturn followed (Tolbert, 1966). Observations of other astronomical sources were made with the 35/70 GHz receiver: Tau A and Sgr A (Tolbert and Straiton, 1965) and the Crab (Tau A) and Orion Nebulae (Tolbert, 1965). An article summarizing observations of the Moon at 35 and 70 GHz over six months of lunar phases (Clardy and Straiton, 1968) was the last publication of Solar System studies with the antenna from the Austin location. Experience had made it clear that a dryer site was required for millimeter wavelength astronomy.

## 3 THE MILLIMETER WAVE OBSERVATORY

It was decided to move the antenna to McDonald Observatory (McD) in West Texas, to a location on the western side of Mt. Locke, at an elevation of 2070 meters. NASA provided funding and the Western Development Labs received a contract for the move. The antenna did not become part of McD, but remained an EERL facility. Straiton and Harlan Smith, McD Director, agreed to this arrangement on Mt. Locke because the antenna would continue to be operated by electrical engineering staff and there was no one in the Department of Astronomy at the time who was interested in millimeter wavelength astronomy. The antenna became a 'satellite observatory' on Mt. Locke, called the Millimeter Wave Observatory (MWO).

A serious problem was encountered in bringing the antenna back into operation. Measurements of the antenna gain showed a substantial loss, which was traced to astigmatism in the beam pattern. The antenna had a polar mount, with the backup structure for the reflecting surface attached to each of the two forks by four bolts. The contractor had either failed to preserve the original shim thicknesses for these bolts or had somehow warped the reflector in transporting it. A long program of beam pattern measurements using a transmitter installed at the nearby Davis Mountains State Park eventually resulted in a properly-shimmed mount and restored antenna gain. (During the restoration of the antenna gain, a Japanese solar astronomer, Wu-Hung Su, used the daytime hours to observe the Sun. The angular size of the Sun meant the broadened antenna beam was of no concern.) The techniques developed in the course of this work, as applied to the MWO antenna, became the PhD thesis of John Davis, supervised by John Cogdell. The work appeared in two publications (Cogdell and Davis, 1973a; 1973b). For an account of the installation at Mt. Locke, a description of the antenna performance, and similar descriptions of the performance of other operating millimeter antennas as the time see Cogdell et al. (1970).

Once the antenna was operational, planetary observations resumed. Although the site was clearly superior to Austin, the poor noise figure of the receivers made for long integration times and the work was slow. Cogdell and Davis, the only observers, could not man the telescope full time; teaching and family duties severely restricted operations. It was during one of these downtimes in the spring of 1971 that two visitors to McD, wandering around the mountain in the late afternoon, happened on the antenna standing idle. One of them, Patrick Thaddeus from Columbia University and the Goddard Institute of Space Studies (GISS), knew something of the antenna, in particular, that it was of high quality. He also knew that Arno Penzias and his group at Bell Telephone Labs (BTL), who had recently discovered

.

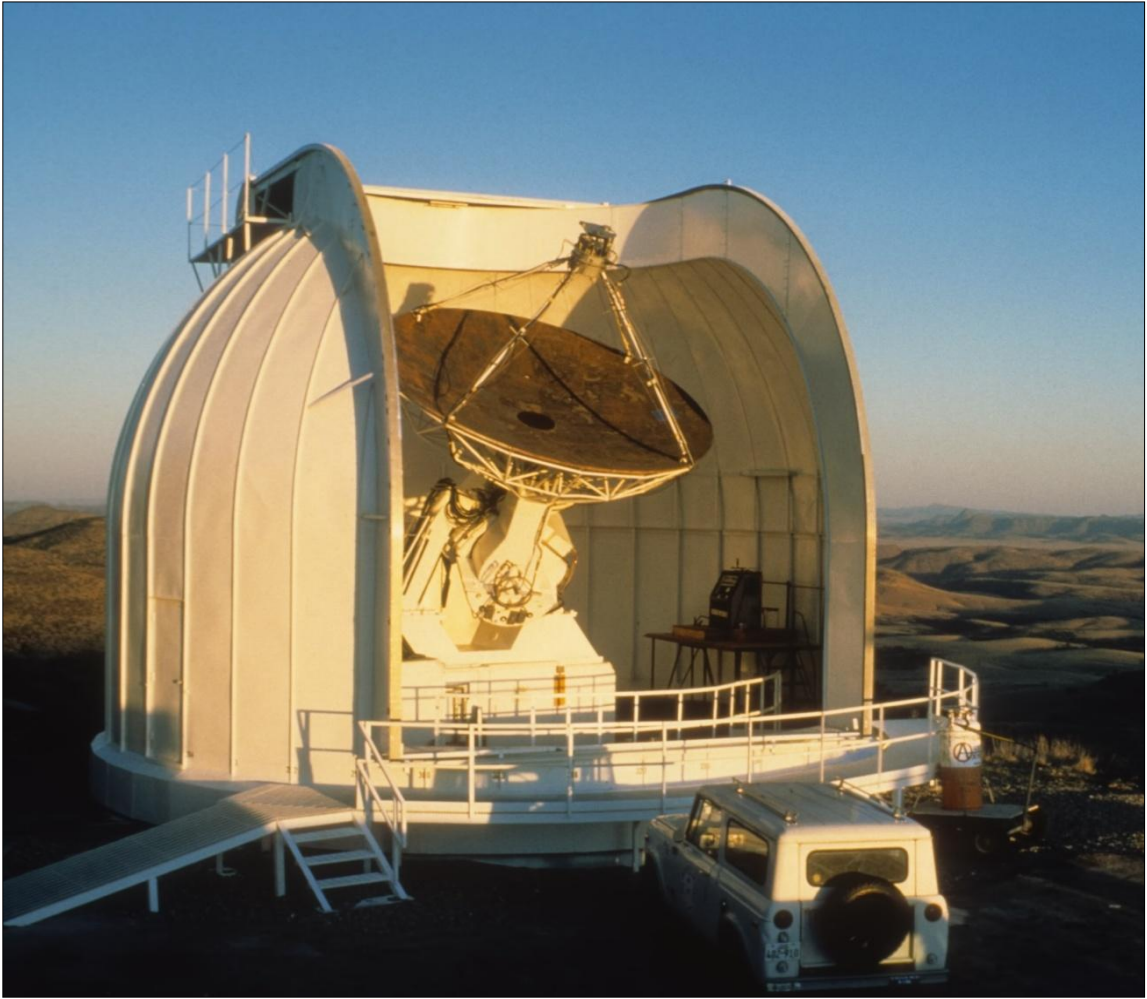

Figure 3: Photograph of the MWO 4.9-meter antenna and astrodome following dome renovations and installation of the error-correcting subreflector and high-frequency receiver box (courtesy: McDonald Observatory).

interstellar CO (Wilson et al., 1970), were unhappy about what they considered the small amount of time they were getting to pursue their discovery at the National Radio Astronomy Observatory's (NRAO) 36-ft Radio Telescope. At the end of their observing run, Thaddeus went home to propose to Penzias that he bring his receiver to the MWO in exchange for observing time. The other observer (P. Vanden Bout) went back to Austin, where he was teaching in the Department of Astronomy, to propose to Cogdell and Davis that they make the MWO available for interstellar molecular astronomy in exchange for using the much superior BTL receiver for their planetary work. The missing element for interstellar molecular spectroscopy was a spectrometer. In due course, Vanden Bout organized a four-way partnership: UT provided the antenna, BTL a receiver, GISS reference oscillators, and the Harvard College Observatory (HCO) a spectrometer. The observing time not required by the EERL for planetary work was split equally between the UT group and each of the other three partners for interstellar spectroscopy. The agreement was merely verbal at the start. It was formalized in writing in 1975, apparently, to satisfy the NSF. The letter, signed by P. Thaddeus, is reproduced in Figure 2.

The first funding for the new project was from McD,

in the amount of $1000, to buy parts for a sidereal clock. Additionally, A. Straiton, who was then serving as Acting Dean of the Graduate School, provided $5000 for the purchase of a spectrum analyzer, vital to the construction and testing of the local oscillator system. The UT group received its first grant from the National Science Foundation (NSF) in 1972. Without the strong support of NSF program officer James Wright, the entire enterprise could have failed. The head of astronomy at NSF regarded the partnership as unmanageable, but at Wright's urging he gave his reluctant approval. NSF funding to UT for the MWO continued until 1988. Wright was the NSF program officer until the early 1980s when Kurt Weiler replaced him. The last NSF program officer was Vernon Pankonin, who took over in 1986. A grant for instrumentation was provided by the Research Corporation, and used to purchase mixer diodes fabricated by Robert Mattauch from the Department of Electrical and Computer Engineering at the University of Virginia. P. Vanden Bout received support from the Welch Foundation.

Observations with the new equipment began in the fall of 1972. Figure 3 shows the antenna and astrodome as equipped in the 1980s. One of the striking features of the antenna can be seen in Figure 3—the surface is gold plated. The thin film of gold was in-tended to protect the thicker silver paint that was the actual conducting surface. Unfortunately, the silver migrated through the gold and tarnished. Although this had no apparent effect on performance, it was thought to be prudent to re-plate the surface from time to time. Obtaining University and federal approval to do this was a challenge and the group endured many jokes about their 'gold-plated' telescope.

The partnership proved very beneficial to the planetary program, which continued to receive strong support from the NASA program officer, William Brunk. Using the new equipment and an improved transmitter and gain calibration system, it was possible to make accurate absolute measurements of planetary brightness temperatures. Results were reported in a series of papers (Ulich, et al. 1973; Ulich 1974; Ulich, et al. 1980) that remained fundamental references for absolute planetary brightness temperatures until supplanted by recent space mission observations. Planetary observations at the MWO stopped in 1977 with the end of NASA funding.

The interstellar spectroscopy programs flourished, enjoying the advantages of good receivers and, compared to what was available elsewhere, large amounts of observing time. The partnership worked remarkably well. Each partner needed the others, and their scientific interests were to a certain degree different. Their sources of funding were also different: NSF funded the EERL to operate the MWO. BTL was supported by AT&T. NASA funded GISS. And HCO had an endowment. At least in the early days, the amiable relations at the MWO stood in contrast to the NRAO 36-ft Radio Telescope, where astronomers engaged in a vigorous competition to gain what was typically a few days of observing time, often to search for a new interstellar molecule. The much smaller MWO telescope lacked the sensitivity easily to discover new molecules; all the strong emitters had already been found. Instead, the observing programs were largely devoted to using the strongest molecular lines to address questions posed by the discovery of an entirely new phase of the interstellar medium, for example, to determine the nature of the molecular clouds and probe their physical conditions.

## 4  THE MWO STAFF

The MWO staff was typical of university research facilities, consisting of faculty and graduate students, supplemented by a minimal number of support positions. After the move of the antenna from Austin to Mt. Locke, responsibility for research and operations moved from Straiton, assisted by Tolbert, to Cogdell and his graduate student, Davis. They, together with another graduate student, Bobby Ulich, conducted the NASA-funded program of measuring planetary brightness temperatures. Responsibility for technical matters remained in Electrical Engineering throughout the history of the MWO. Davis was a research associate after graduation in 1970 and joined the faculty in 1978. His graduate students, Charles Mayer and Heinrich Foltz, made major contributions to holographic antenna evaluation. The Department of Electrical Engineering was a source of skilled labor: graduate students Natalino Camileri and Rod-ney Barto assembled a receiver and a paged memory system for the NOVA computers, respectively. Wan Ho, another electrical engineering student, also worked on receivers. Wolf Vogel took over the propagation studies research program at EERL in 1969, and was a valuable resource for technical matters at the MWO. Vogel was honored as a Fellow of the IEEE for his work on propagation modeling in 1991. William Wilson worked with the Texas group as a faculty member in Electrical Engineering during 1976-1977. Anthony Edridge was a receiver engineer in 1984-1985 and he built a 2mm receiver, and as a postdoc Steve Laycock worked on a 350 GHz receiver.

The Department of Astronomy at UT gave the MWO research program a major boost in 1975 with the appointment to the faculty of Neal Evans, who introduced an emphasis on the physics of molecular clouds to the research program. Frank Bash moved to the MWO group in 1980, following the completion of the operational phase of the Texas All-Sky Survey, which had been conducted near Marfa, Texas, by the Jim Douglas group. Bash had responsibility for the MWO in its final years of operation. Another Department of Astronomy faculty member, Dan Jaffe, was part of the group in 1988. His experience in submillimeter astronomy was helpful in developing the possibilities for converting the MWO antenna to a submillimeter facility. Herbert Pickett, a molecular spectroscopist in the Department of Chemistry, was associated with the MWO in the mid-1970s. Other UT faculty that encouraged the development of the MWO included James Browne of the Department of Computer Engineering and James Boggs of the Department of Chemistry, who were theoretical and experimental molecular spectroscopists, respectively. They helped launch a series of seminars to introduce MWO group members to molecular quantum structure.

The support staff was a part of the EERL. Wanda Turner handled all administrative affairs from basic secretarial support to purchasing and accounting, working at the EERL from the days of Straiton to the closure of the MWO. A.J. Walker provided mechanical engineering support as the group machinist over the same period. He was critical to the maintenance of the radio telescope and dome as well as all mechanical components of the MWO. Charles McEvoy served as an electronics technician at the EERL, building professional quality electronics for the MWO for several years. The remote location of the telescope required on-site care. Carlos Garza, an electronics technician with multiple skills, provided that for many years. Garza was a Navy veteran, where he had been a radar technician. Larry Strom succeeded Garza. Strom had had a TV cable installation business in Dallas, Texas. On closure of the MWO, Strom became the on-site technician for the Caltech Submillimeter Observatory in Hawaii. Loren joined the MWO staff on completion of his graduate research. After aiding in the construction of a filter bank in the summer of 1977 and taking it to the telescope for installation, he remained there, becoming, in time, the 'man on the mountain', a 'go-to person' for all observers, as well as a key contributor to MWO science.

Critical technical support came from the partners, in particular, Robert Wilson (BTL), Keith Jefferts (BTL), Anthony Kerr (GISS), and Hays Penfield (HCO). Guest observers who brought receivers to the MWO included Glenn White (Queen Mary College, London), Dick Plambeck and Paul Goldsmith (University of California, Berkeley), and Tom Phillips (BTL). Significant technical help with receivers was provided by Neal Erickson (Five Colleges Radio Astronomical Observatory, University of Massachusetts) and with local oscillators by John Payne (NRAO) and John Carlstrom (Caltech).

It is a sad but curious fact that the incidence of Parkinson's Disease appears to be anomalously high among senior radio astronomers. Overall, Parkinson's Disease strikes one in a thousand men over the age of 70 (Van Den Eeden, et al., 2003). At least ten older radio astronomers out of a population that cannot exceed roughly one hundred suffer from Parkinson's. To date, this includes only one of the observers at the MWO. If there is an environmental factor at work here, it may lie in the laboratory, where some pioneering interstellar spectroscopists spent much of their time, rather than at the telescope.

# 5 THE SCIENTIFIC PROGRAM AND INSTRUMENTATION

## 5.1 The Early Days

As with any telescope, the research is limited by the instrumentation. The MWO antenna had an aperture of 4.9 m, giving it a beam size (FWHM) of 2.6′ at the frequency of the CO (J=1-0) transition (115 GHz), by far the most heavily-observed molecular line. It operated at prime focus, limiting the physical size of receivers. Its surface accuracy of 90 μm rms provided good efficiency, and it enjoyed extraordinary thermal stability, due to the construction of its surface panels and backup structure in Invar, a very low thermal expansion metal alloy. Observing with the surface half illuminated by the Sun made no measureable difference in the antenna gain. The absolute pointing accuracy was 22″ and the tracking accuracy was 7″. An astrodome provided protection from the wind and weather, but it had to be closed in winds over 56 km/h or risk the doors being lifted out of their tracks. This limit was tested by Bobby Ulich, who, ignoring the rule, observed on a windy day and had one of doors unseated by an 80 km/hr gust. Initially, the rotation of the dome was under manual control.

The drive system used opposing torque motors on both the polar and declination axes, with position read by shaft encoders. To track a source, one drove the telescope under manual control to the position the source would have at an upcoming even minute of time. At the right instant, one pushed a switch to start the tracking. Ephemerides had to be printed out in advance for every object to be observed. Because the antenna had been built to track Solar System objects, the clock ran on ordinary solar time. One of the first improvements was to build a sidereal clock. The absolute rms pointing accuracy was within a quarter beam width provided one had taken care to measure offsets in Right Ascension and Declination by looking at one or two planets that might be visible.

The Bell Labs receiver was built out of waveguide and used a Schottky-barrier diode mixer. The diodes were mounted in Sharpless wafers and the receiver was tuned using three micrometers, two for the cavity ends, and one for the backshort. Local oscillator (LO) power came from klystrons. Tuning the receiver was a long stroll through a multi-parameter space, adjusting the micrometers and LO power, measuring the response to hot and cold loads, and repeating the procedure until the maximum response had been achieved. By modern standards, the receiver was primitive, but for its day it defined the state of the art. Typical double-sideband noise temperature on the sky was ~1000 K. Their unique feature was the use of wafers that held the smallest area Schottky-barrier diodes available at the time.

The reflex klystrons for the LO operated at high voltage, had limited lifetimes, produced limited power, and were very expensive. A division of Varian Corp. in Canada produced the millimeter frequency units that were required. The market was small and eventually loss of key personnel at Varian meant no klystron could be purchased for use at frequencies above 100 GHz that would actually work. The supply of klystrons in hand carried the MWO to the end of its operation. Alternatives to klystrons were considered. In particular, backward wave oscillators known as Carcinotrons, which produced significant power up to THz frequencies. Thijs de Graauw brought a Carcintron to the MWO for a test, but the cost of these systems precluded their routine use at the MWO.

All LO chains were phase-locked to a frequency reference. At first, the MWO used a tunable frequency synthesizer to generate a reference signal near 100 MHz, which was then fed into a resonant cavity to reject all but the 20th harmonic. In turn, a signal typically lying between the 55th to 75th harmonic of the 2 GHz signal was used to phase-lock the klystron. The usual observing mode used frequency switching, accomplished by feeding the klystron phase locking circuit with two alternating reference signals, separated by 80 MHz. Calculating the correct observing frequency to account for the source velocity in the local standard of rest and the motion of the telescope on the surface of the Earth was done by hand. It did not take long to acquire the computer program used at NRAO. Similarly, if power was lost and the sidereal clock needed to be reset, a hand calculation using the U.S. Naval Almanac was required.

An L-band parametric amplifier was used to amplify the intermediate frequency. It operated at 1.4 GHz with a bandwidth of 100 MHz. So-called paramps were the curse of radio astronomy until cooled-transistor amplifiers replaced them. The problem was gain stability. Some progress was made at the MWO by mounting the paramp to a water-cooled metal plate, limiting thermal drift.

The first spectrometer, built by Hays Penfield of HCO, had two sets of (40) channels, one of width and spacing equal to 250 kHz and one to 2 MHz. The voltages on the capacitors that accumulated the signal were read out digitally (and very slowly).

Figure 4 shows an early CO dark cloud spectrum.

The data were recorded on punched paper tape for offline data reduction. For several years, Amber Woodman of McD made plots of spectra from punched paper tapes on an x-y plotter located in the 107-inch Telescope dome. The software that produced the paper tapes was written by Bob Wilson and called 'BTL'. To initiate an observation, one pressed the Spectral Line Observe (SPLOBS) button. This needed to be done for each individual integration. Failure promptly to start a new integration wasted observing time and P. Thaddeus posted a sign reading "DON'T THINK - INTEGRATE". In time, a thumbwheel was added to the SPLOBS button that set the number of integrations desired. Using a high number kept things going, and integrations were only stopped for emergencies: grass fires that threatened the facility, a tropical storm remnant that took out the power lines in flash flooding, and the all-too-frequent lightning strikes in summer.

The observed line intensities were calibrated using a chopper wheel technique developed by the BTL group (Penzias and Burrus, 1973). The system produced a calibration signal proportional to the difference between an ambient temperature mm absorber and the sky. Conveniently, the technique did not require knowing the atmospheric opacity. The technique was refined in an early paper by the Texas group (Davis and Vanden Bout, 1973) to take account of differences in gain and atmospheric absorption between the two sidebands of the receiver. For molecular clouds observed at low elevation and for certain receiver setups these effects could be significant.

The inefficiencies and limitations of the equipment were compensated for by the strength of the CO lines, the relatively large amounts of observing time available, and the high quality of the receiver. The MWO was a CO-mapping machine, most particularly in its early days of molecular line observing, which began in the fall of 1972. The antenna was just the right size, large enough to identify the locations of CO hot spots, where the very youngest stars are embedded, but small enough to map a galactic cloud in a reasonable time.[1]

The first publications reported mapping CO in nebulosity associated with Herbig Be/Ae stars (Loren et al., 1973) and in the Orion Nebula (Tucker et al., 1973). An exception for that first observing season was the detection of a new transition in SO (Gottlieb and Ball, 1973) by members of the Harvard group, who had discovered interstellar SO with the NRAO 36-ft Radio Telescope. Overall, approximately two-thirds of the observing time was devoted to CO, the rest to studies of other molecules. A CO map of L43, shown in Figure 5, was typical of those made in the early years (Elmegreen and Elmegreen, 1979).

### 5.2 The MWO Reaches Maturity

#### 5.2.1 Facility Development

The success of the MWO in its early years, and the demands of heavy observing schedules for more reliability, led to a program of improvements that was carried out over a number of years. Central to this program was the construction of a small building adjacent to the telescope dome that could house the control system and all electronics aside from the receiver. The old control room was located in the dome behind the telescope support piers. It was extremely crowded; no more than three people could fit inside at any one time. Figure 6 shows a student observer, E. (Betsy) Green, in the new control room. UT provided the funds, through McDonald Observatory. Support from the Vice-President for Research, Gerhard Fonken, was the key to getting the building and to making improvements in the dome itself. Archie Straiton wanted to use the dedication of the building as an excuse to get a distinguished Texan to visit the MWO. He had Lady Bird Johnson in mind. But Harlan Smith, McD Director, felt that Lady Bird deserved to cut the ribbon for something bigger than the little MWO electronics house, and after an exchange of memoranda the idea was dropped. It is ironic that Lady Bird had previously enjoyed a visit to the still smaller transmitter hut in the Davis Mountains State Park that was used to map the antenna pattern.

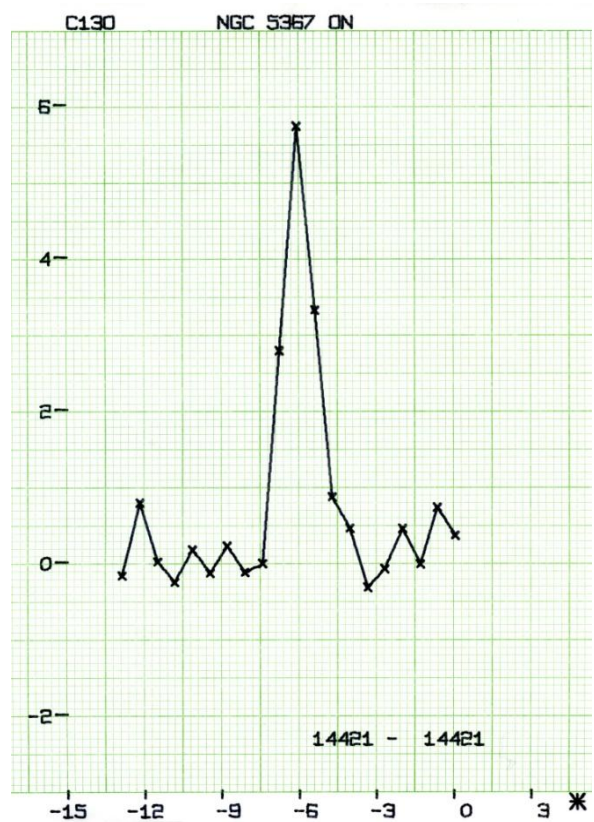

Figure 4: A $^{13}$CO (J=1-0) spectrum in NGC5367, as plotted on the x-y plotter in the McD 107-inch Telescope control room. This was the only data display available to the observer during an observing run in the early years (from the personal papers of R. Loren).

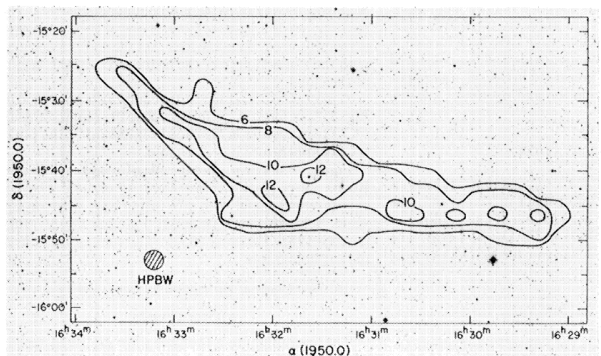

Figure 5: Map of the dark cloud L43 made in the CO J=1-0 transition. The CO contours are superimposed on the red plate of the Palomar Sky Survey. (after Elmegreen and Elmegreen, 1979. © American Astronomical Society. Reproduced by permission).

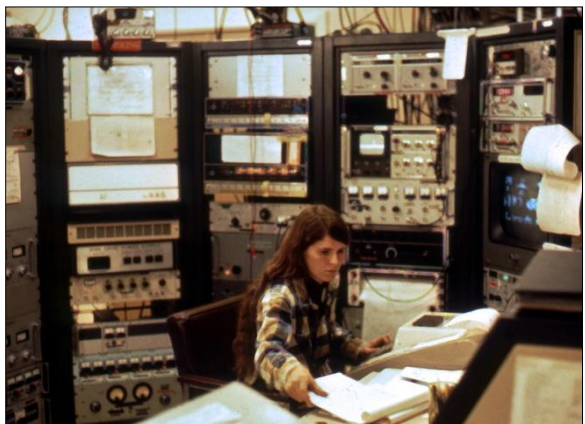

Figure 6: Photograph of the new MWO control room. The student observer is Betsy Green (from the personal papers of R. Loren).

Replacing the wing doors to the dome with a roll-up door allowed the wind limit to be raised to 88 km/hr. This significantly increased the amount of observing time on the windy MWO site. Spring brought high winds, with gusts recorded as high as 160 km/hr. One such gust brought down the power lines. The observer, L. Mundy, and site technician, L. Strom, were forced to turn the dome to face into the wind by hand, using a so-called 'come along', a device for making barbed wire fencing taut.

During this period the entire electronics system was replaced and the software upgraded. The filter bank was expanded to 256 channels each of 62.5 kHz and 250 kHz, and eventually 512 channels of 1 MHz. The new filter bank was pipelined (with what would be called an embedded processor today), so that while all of the channels were being read a new integration was under way. The software upgrade consisted of automating several functions that had previously been handled manually by the observer, such as the dome positioning, and keeping the antenna within its pointing limits.

In contrast to major optical and radio telescopes which were operated by night assistants and telescope operators, the MWO was a strictly do-it-yourself facility. The observer alone did all the operations required. For this mode of operation to be successful required a very user-friendly computer program, at a time when computers were not all that friendly. Bill Peters and John Davis designed the computer software and hardware interfaces. The control program was called NIMBUS and ran on a single NOVA computer. It handled all telescope and dome functions, controlled the receiver frequency, and acquired the data. Another program called ABACUS, running on a second NOVA computer, was used for data reduction. The two computers shared a single hard drive. These computer programs were notable for their reliability and simplicity of operation. After a brief introduction, first-time novice student observers could take full control of the telescope, acquire data, and reduce it while taking more data.

The MWO antenna drive featured two rate zones to optimize observing time. A very rapid slew rate was used when the antenna position was changed by a large amount. As the final source position was approached, the drive shifted into a critically-damped regime to arrive at the source at the exact tracking velocity. The antenna drive consisted of software embedded in NIMBUS that drove linear amplifiers directly connected to opposing DC torque motors.

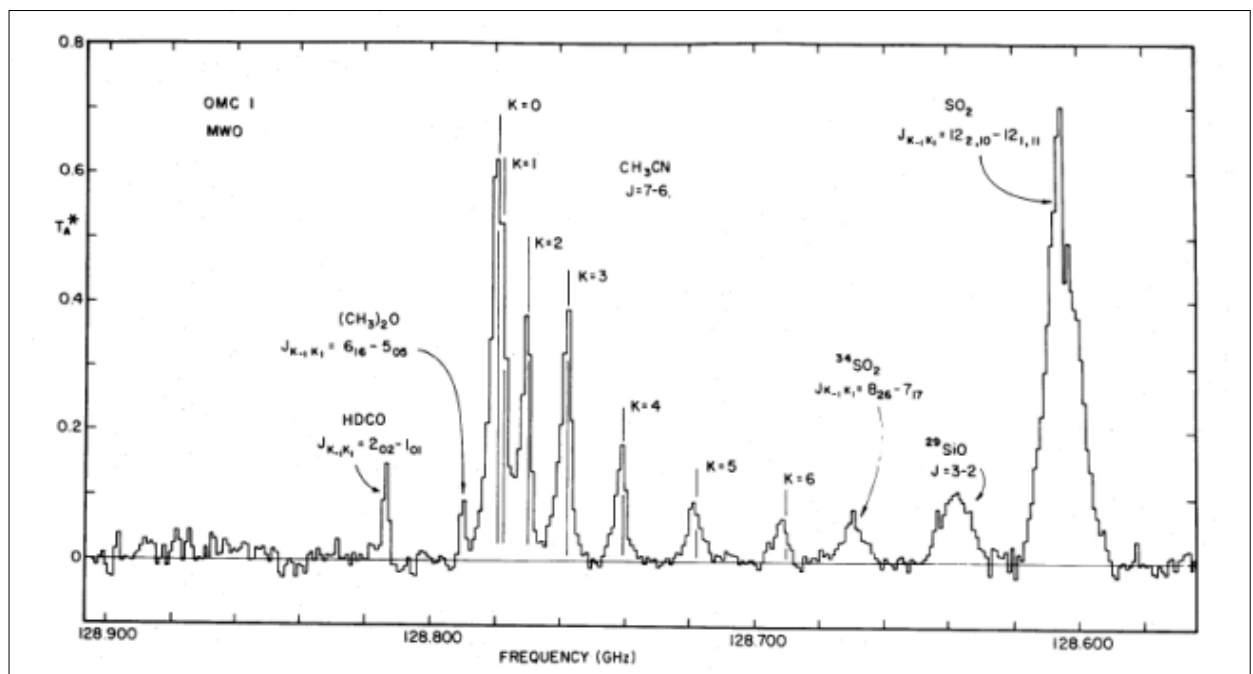

Figure 7: Spectrum of the Orion molecular cloud in the vicinity of 128 GHz, showing the K-ladder lines of $CH_3CN$ (J=7-6), and lines of $SO_2$, SiO, $(CH_3)_2O$, and HDCO (after Loren and Mundy, 1984. © American Astronomical Society. Reproduced by permission).

The priorities for science-related improvements were always higher than those for creature comforts. For example, the two mobile homes used for observer lodging, obtained as surplus government property, were never kept at anything more than the bare minimum needed for shelter. One spring a wind gust of 160 km/hr removed the shade roofing over the trailers and wrapped it around the power lines. On other occasions, observers were forced to nail down loose siding to prevent it being torn off in a storm. Mt. Locke was plagued with insects. No observer could forget the seasonal invasion of moths, let alone the spiders and scorpions. Some observers refused to use the poorly-sealed and poorly-insulated mobile home accommodations and stayed instead in the McD Transient Quarters. Others found solace in a diet of steak and beer, supplemented by Cuban cigars that were occasionally brought in by French observers (see Section 7). The ban on alcohol in UT facilities was widely ignored at McD, which was over 600 km from the University in Austin.

Simply travelling to the MWO was difficult, due to its remote location. Some observers drove from the nearest airport, either Midland/Odessa (250 km) or El Paso (320 km). Others drove all the way from Austin. Travelers always checked to see if there was a spare seat on the plane chartered by McD for weekly trips between Austin and Marfa, a town 50 km from the Observatory with a landing strip that had been used to train WWII glider pilots. The most economical mode of travel was to take the Greyhound bus to either Alpine or Kent. The latter was no more than a very isolated gas station in the tumbleweeds off the highway between Houston and El Paso. Someone from the MWO would pick up and drop off travelers at these spots. Exiting a bus after the gas station in Kent had closed for the night was done with trepidation if the MWO driver was late; one waited in the dark with only the howling of the wind and coyotes for company.

### 5.2.2 Scope of the Research

The MWO produced over 250 papers published in refereed journals and conference proceedings, as well as 23 PhD dissertations. Only a small fraction of these can be discussed here. There was a wide range of topics, illustrated by these papers: the interaction of supernova remnants with molecular clouds (Wootten, 1977; 1981); the enormous extent of molecular cloud complexes (Elmegreen and Lada, 1976; Kutner et al., 1977; and Lada et al., 1978); molecular clouds associated with giant HII regions (Lada, 1976); the dynamics of CO clouds and galactic density waves (Bash and Peters, 1976; Bash et al., 1977); circumstellar shells (Lambert and Vanden Bout, 1978; Clegg and Wootten, 1980; Sahai et al., 1984); a limit on the abundance of oxygen in molecular clouds (Liszt and Vanden Bout, 1984); high galactic latitude molecular clouds (Blitz et al., 1984; Magnani et al., 1985); CS (J=5-4) line profiles towards Sgr A at 245 GHz (Sandqvist, 1989); molecular rotational constants of CCH (Ziurys et al., 1982); and molecules in comets (Irvine et al., 1984). Figure 7 shows a small portion of the rich molecular spectrum from the Orion molecular cloud, obtained in a study of $CH_3CN$ (Loren and Mundy, 1984).

In the Texas group, star formation was a topic of central interest. To understand star formation, one needed to know the physical conditions of the dense molecular cloud cores where stars formed. Neal Evans defined a research program that continued for many years to probe the physical conditions of cloud cores. In a series of papers, the energetics of molecular clouds was examined (Evans et al., 1977, 1981; 1982; Blair et al., 1978; Evans and Blair, 1981), comparing the energy input from newly formed stars with cooling by molecular line emission and far-infrared emission by dust. The optically thick CO line gave the gas kinetic temperature. Cloud cores were located by mapping $^{13}CO$, which peaked in emission strength on the densest gas and also gave a rough location of potential-embedded stars. Near infrared observations were made to characterize the stars, and published far-IR data were used to determine the dust luminosity. Observations of both the 2 mm and 2 cm formaldehyde lines at the MWO and with the NRAO 140-ft Radio Telescope, respectively, gave a good estimate of core densities. The utility of CS and $H_2CO$ as density tracers for modeling clouds was demonstrated in a series of papers (Snell et al., 1984; Mundy et al., 1986; 1987).

### 5.2.3 Toward Submillimeter Observing, the 1mm Band

At the outset, observations at the MWO were limited to frequencies between 110 and 150 GHz. But it was clear that a niche for the MWO was the 1mm frequency band. Higher frequencies were unexplored, as receivers at these frequencies did not exist. The first experience at higher frequencies at the MWO was with receiver components developed by the innovative radio astronomy group at the University of California, Berkeley. The receiver was tuned to respond to the second harmonic of the LO reference signal. In 1979, Richard Plambeck's diplexer mixer was used to make the first observations of the J=2-1 transition of CO in a bipolar outflow source, showing that temperatures in the outflow were up to a factor of three hotter than the surrounding cloud (Snell et al., 1980). It was also used to detect the $3_{12}$-$2_{11}$ line of $H_2CO$ at 1.3 mm (Evans et al., 1979).

Experience with so-called second harmonic receivers prompted the Texas group to examine the effect of second harmonic response in a standard receiver tuned to the fundamental of the LO (Vanden Bout et al., 1985). Because the effect was usually small, observers tended to ignore it. Everyone in mm astronomy at the time knew and accepted the seemingly unavoidable systematic uncertainties in calibration. In time, the advent of mm wavelength interferometers and single-sideband receivers allowed line calibration to be more precise. The Atacama Large Millimeter/Submillimeter Array promises an intensity accuracy of 5% in line images.

The University of Massachusetts' N. Erickson (1977) from the Five Colleges Radio Astronomy Observatory built the first quasi-optical receiver using a Martin-Puplett interferometer for local oscillator injection and a clever refocusing of the prime focus spot to form the quasi-optical beam. For a description of this system see Goldsmith (1988). Producing a local oscillator signal for frequencies above 150 GHz was a challenge, accomplished by using the 2nd, 3rd, or even 4th harmonic of a klystron tuned from 80 to 115 GHz. Subsequently, J. Davis and C. Mayer mapped the antenna surface error pattern with a holography receiver they developed (Mayer et. al., 1983). They made a folded Gregorian optical system with an error-correcting secondary mirror, which was installed into the beam path in 1983. This optical system is shown in Figure 8. The error corrector significantly improved the antenna efficiency in the 1mm band. An illustration of the power of this new receiver/optics system is the study Mangum et al. (1990) made of $H_2CO$ in OMC-1 (all 14 $\Delta J$=1 lines of $H_2CO$ between 211 and 363 GHz, 8 lines of $H_2^{13}CO$, and 4 lines of $H_2C^{18}O$). The higher excitation of lines from levels in the K=2 and K=3 ladders, which only occur at higher frequencies, are especially useful for determining the physical conditions in the hot core and plateau regions of the cloud. They combined these data with VLA observations to determine the physical distribution of $H_2CO$ in the kinematic components of OMC-1.

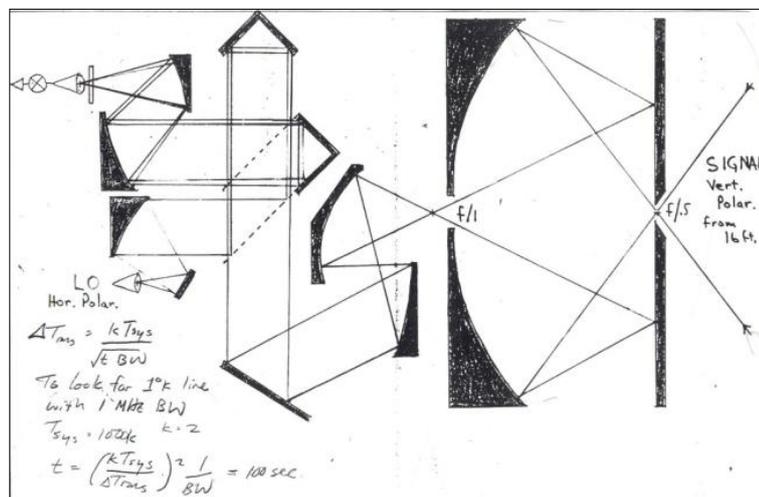

Figure 8: Drawing by Mayer and Davis of the folded error-correcting secondary and associated receiver optics. The errors in the primary surface were cut (with the opposite sign) into the mirror just behind the f/0.5 primary focus. The mirror is smaller than the primary by a factor of ~70, and was made on a numerically-controlled milling machine driven by a program derived from the holography map (see Figure 10). Sketches such as this were typical of documents used in construction of equipment; the interactions with technicians and machinists were close and informal (drawing from the files of the EERL).

In 1985 observations were pushed to true sub-millimeter wavelengths (301–352 GHz) with cooled Schottky diode mixers developed in-house by N. Camilleri. The receiver used Erickson's quasi-optical injection system, with tripled and quadrupled klystron fundamental frequencies for the LO. Over twenty-four new lines were detected in a host of molecules: $H_2CO$, SO, $SO_2$, CS, $C^{34}S$, $CH_3OH$, CN, CCH, SiO, and $H^{13}CN$, all in OMC-1 (Loren and Wootten, 1986).

Studies of deuterated molecules, a program of interest to the Texas group, were aided by the high frequency capability. In contrast to atomic deuterium, which has only recently been detected in the inter-stellar medium of our Galaxy, deuterated molecules are relatively easy to detect. Snell and Wootten (1977) detected DNC. Combes et al. (1985) detected CCD. Using an InSb bolometer receiver, Beckman et al. (1982) detected the $2_{11}$–$2_{12}$ line of HDO at 242 GHz and estimated the $HDO/H_2O$ abundance ratio. Molecular ions like $HCO^+$ recombine with free electrons. Guelin et al. (1977) and Wootten et al. (1979) used the $DCO^+/HCO^+$ ratio to find upper limits for $X_e$ of order $10^{-7}$ to $10^{-8}$ in a number of Galactic dark clouds. Wootten et al. (1982) confirmed strong temperature dependence for $DCO^+$ fractionation in a wide range of clouds, obtaining results similar to those of Snell and Wootten (1979) for DNC. $DCO^+$ became a marker for cold, star-forming cloud cores. Loren et al. (1990) mapped $DCO^+$ emission in a cluster of twelve such cores in the ρ Ophiuchus molecular cloud. The cores contained infrared sources having steep spectra characteristic of the youngest protostars, indicating a brief phase of evolution during which $DCO^+$ molecules exist before stellar heating destroys them.

### 5.3 Significant Results and Discoveries

The principal advantage of the MWO was that it had sufficient observing time to allow for extensive map-ping of molecular emission. This led to what are its most highly cited results: the discovery of mass out-flows from newly-formed stars; a means of measuring molecular cloud mass; evidence for the dark matter halo of our Galaxy; and the discovery of an interstellar maser. The most significant technical achievement at the MWO was the development of holographic antenna evaluation.

### 5.3.1 Bipolar Outflows

It was widely assumed in the early years of milli-meterwavelength astronomy that star formation occurred in molecular clouds, and much research was devoted to that connection. Looking for kinematic evidence for the collapse of molecular gas onto a new star was an early interest of Bob Loren. Among the first clues for collapse was the observation of broad CO line wings in the cores of the Mon R2, R CrA, and LkHα198 clouds (Loren et al., 1974). Mon R2 was particularly intriguing (Loren, 1977). It's CO line shows a self-absorption feature shifted by 1 km/s with respect to the $^{13}CO$ line, indicating motion of cold outer gas toward the center of the cloud. The CO lines showed a bipolar structure that was aligned with the rotation axis determined from $^{13}CO$ mapping. The conclusion of the paper, reflected in its title, was that the flow was inward, but the data in hand could not rule out an outward flow in the bipolar structure.

That bipolar flows were outward was established from observations of the cloud L1551, published in what is now seen as an iconic paper (Snell et al., 1980). Their CO map, reproduced in Figure 9, shows red- and blue-shifted emission extending 0.5 pc in both directions from a central star, IRS-5, visible only in the infrared. The blue-shifted lobe contains knots of nebulosity, HH28, HH29, and HH102, whose optical radial velocities and proper motions (Strom et al., 1974; Cudworth and Herbig, 1979) yield a true space motion away from IRS-5. Identifying these knots as associated with the blue lobe implies outflow from the star. The model presented in their paper has come to be the standard picture of a stage in star formation when the collapse of the molecular cloud core has formed a rotating protostellar disk with a stellar wind/shock wave that flows out along the rotation axis. These outflows are ubiquitous signatures of newly-formed (low-mass) stars and have been studied in great detail (see the review by Lada, 1985). Bipolar outflows are arguably the most significant discovery made at the MWO.

### 5.3.2  CO to Molecular Cloud Mass

The hydrogen molecule, principal cloud constituent, is not easily observed and the mass of molecular clouds is typically estimated using observations of CO, which is easily detected at millimeter wavelengths. The X-factor converts the brightness temperature of the CO emission line to a column density of $H_2$. The original determination of $X(CO)$ began with work done at the MWO by Bob Dickman, a student of Thaddeus, who compared the strength of the $^{13}CO$ (J=1-0) line in so-called 'dark clouds' to their visual extinction. He inferred the amount of dust along the line of sight from the reddening, and, in turn, calculated the column density of $H_2$ using the dust to $H_2$ ratio from independent UV satellite measurements. Because a cloud's mass is a fundamentally important property, this paper (Dickman 1978) is one of the most significant results from the MWO.

### 5.3.3  Kinematics and Dynamics of the Milky Way

The discovery of interstellar CO provided a new means for mapping the structure of our Galaxy. Early work by Leo Blitz (1979) at the MWO showed the utility of CO for such work by obtaining a rotation curve for the outer Galaxy from observations of HII regions in the second and third quadrants. He showed that our Galaxy has a flat rotation curve to a large distance from the Galactic Center, as had been seen in external galaxies from HI observations. This result is among the earliest evidence for our Galaxy's dark matter halo. As an aid to further work, he published a catalog of CO observations towards essentially all optically visible HII regions in the Milky Way. The catalog included nearly all the HII regions in the Sharpless Catalog as well as 65 additional HII regions. This catalog (Blitz et al., 1982) has proven to be of high utility.

### 5.3.4  SiO Maser Lines

Early results from the MWO included detections of rotational transitions of vibrationally-excited (v=1) SiO. Snyder and Buhl (1974) had seen a set of unidentified, narrow lines in the Orion molecular cloud and suggested that they could be the J=2-1, v=1 SiO line. The Texas/GISS groups then observed the v=1, J=3-2 SiO line at the same velocity (Davis et al., 1974). They then detected the v=1, J=1-0 line at this velocity (Thaddeus et al., 1974) using a receiver specially built for the purpose. These observations confirmed the suggestion of Snyder and Buhl, and established the existence of a new interstellar molecular maser.

### 5.3.5  Development of Holographic Antenna Evaluation

The 1970s saw the transition from mechanical to electronic techniques for mapping the surface accuracy of reflector antennas. Scott and Ryle (1977) were the first to use the so-called 'holographic technique' for an interferometer. Bennett et al. (1976) demonstrated the technique on a single dish of 3 m diameter, making partial surface maps that established a proof of concept.

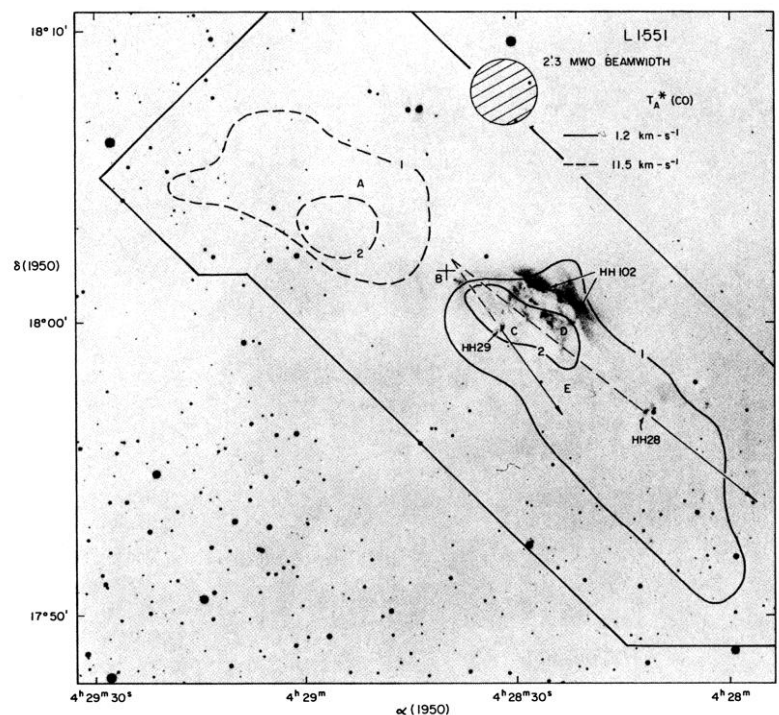

Figure 9: Contours of the CO emission in L1551 superimposed on an optical image, showing the blue and red shifted lobes to the SW and NE, respectively. The blue lobe engulfs the HH objects (after Snell, Loren, and Plambeck, 1980. © American Astronomical Society. Reproduced by permission).

To the best knowledge of the authors, the first holographic map of a large aperture reflector surface, done with sufficient accuracy and resolution to be useful, was that made by Mayer et al. (1983) of the MWO. The surface map is shown in Figure 10. The results of their holographic receiver system were used to design the error-correcting optics that greatly enhanced the MWO performance at 1mm wavelength. Collaboration with NRAO led to the installation of a similar system on the NRAO 12-m Radio Telescope. Holographic evaluation of antenna surface accuracy has become the standard technique in use today.

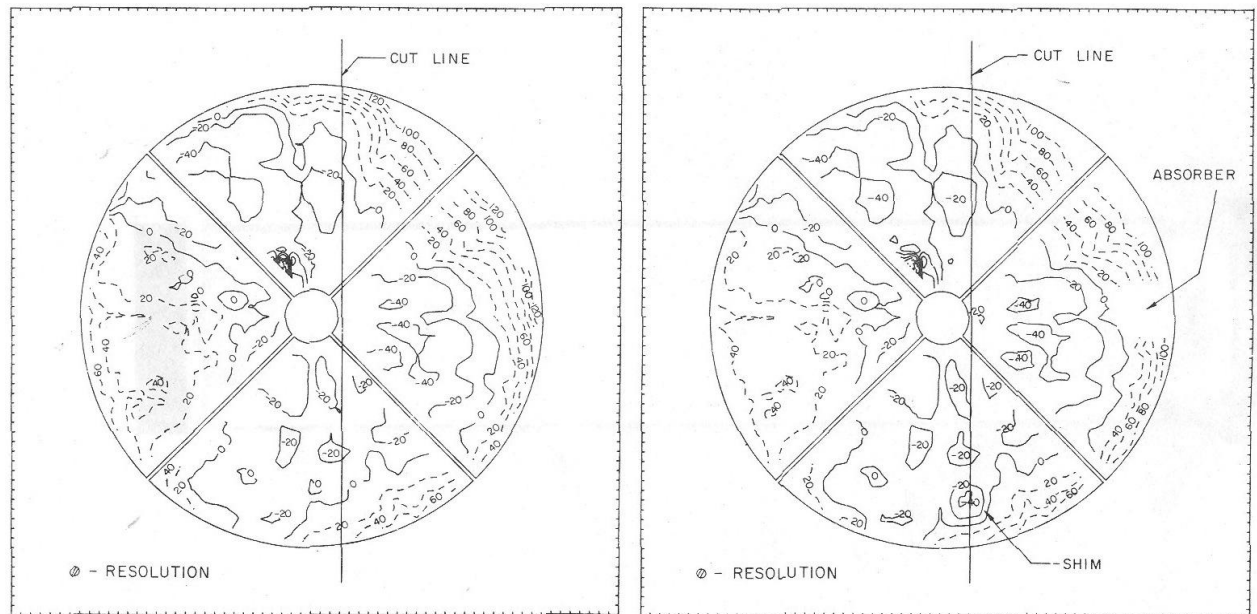

Figure 10: Holographic maps of errors in the surface of the 4.9 m MWO antenna. Left: The normal surface. Right: A sheet of absorber placed on the surface can be seen as an absence of contours. A metal shim taped to the surface is easily seen and serves to confirm the sign of the surface errors with respect to a perfect parabola (after Mayer, et al. 1983. © IEEE. Reproduced by permission).

## 6  STUDENTS, POSTDOCTORALS, AND SABBATICAL VISITORS

Students played a major role at the MWO, where sufficient observing time was available to do large PhD projects. It also contributed to the training of many other students who came to observe for research projects outside their PhD. A list of students who observed at the MWO, with titles of 23 PhD dissertations that included MWO data is given here:

Greg Baran, Guy Blair ("Millimeter Molecular Line and Infrared Observations of Dense Clouds Associated with Small Hα Emission Regions"), Leo Blitz, Elizabeth Bozyan, Ron Buta, Harold Butner ("Dense Cores and Young Stellar Objects"), John Caldwell, John Carr, Fabienne Casoli, David Chance, Gordon Chin, Françoise Combes ("Dynamics and Structure of Galaxies"), Dan Clemens, Hong-Ih Cong, Jacques Crovisier ("Contribution to the Study of the Interstellar Medium by Observation of the 21-cm of Neutral Hydrogen in Absorption"), John Davis ("The Evaluation of Reflector Antennas"), Robert Dickman ("The Ratio of Carbon Monoxide to Molecular Hydrogen in Interstellar Dark Clouds"), Debra Elmegreen, Robin Frost, D. Garrett, Maryvonne Gerin ("Molecular Clouds and Dynamics of Interacting Galaxies"),[2] David Gilden, Betsy Green, Stephane Guilloteau, Paul Ho, John Howe, Frank Israel, Marshall Joy, Charlie Lada ("Observations of Dense Molecular Clouds"), Elizabeth Lada ("Global Star Formation in the L1630 Molecular Cloud"), David Leisawitz, Emmanuel Lellouch, Russell Levreault ("Molecular Outflows and Mass Loss in Pre-Main-Sequence Stars"), Harvey Liszt ("Carbon Monoxide Studies of Hydrogen-II Regions"), Bob Loren ("Millimeter Wavelength Molecular Emission Associated with the Massive Young Herbig Be and Ae Stars"), Robert Lucas ("Study of the Formation of Millimeter Molecular Lines in Interstellar Clouds"), Paul Makinen, Loris Magnani ("Molecular Clouds at High Galactic Latitudes"), Jeff Mangum ("The Throes of Star Formation"), Charlie Mayer ("Microwave Antenna Metrology by Holographic Means"), Marshall McCall, Lee Mundy ("The Density and Molecular Column Density Structure of Three Molecular Cloud Cores"), Anneila Sargent ("Molecular Clouds and Star Formation"), Michael Scholtes, David Slavsky, Ron Snell ("A Study of Interstellar Dark Clouds"), Bobby Ulich ("Absolute Brightness Temperature Measurements at Millimeter Wavelengths"), Peter Wannier ("Isotopic Abundances in Interstellar Clouds"), Bruce Wilking, Diane Wooden, Al Wootten ("A Study of Molecular Clouds Near Supernova Remnants") and Shu-Dong Zhou ("Small Scale Structures and Density of Star-Forming Regions").

The MWO had a good number of observers who came as postdoctorals. Their motivation often went beyond the obtaining of research data to include practice and training in making observations at millimeter wavelengths. Postdoctoral visitors included: John Beckman, John Black, François Boulanger, Jorge Canto, Bruce Elmegreen, Pierre Encrenaz, Steve Federmann, Edith Falgarone, Carl Gottlieb, Elaine Gottlieb, Thijs de Graauw, Michel Guelin, Marc Kutner, Richard Linke, Gillian Knapp, John Mather, Mark Morris, Phillip Myers, Antonella Nata, Peter Phillips, Jean-Loup Pujet, Mark Reid, Luis Rodriguez, Nick Scoville, and Ken Tucker. Howard Van Till was a visitor for the 1974-1975 academic year. He came on an NSF-funded sabbatical to gain experience in astronomical research in support of his new program in astronomy at Calvin College.

## 7  INTERNATIONAL OBSERVERS

The MWO had a significant number of international observers. The largest contingent by far was from France, as can be seen from the list of students given above. It began when Pierre Encrenaz took a postdoctoral position at GISS with Thaddeus, funded by the National Research Council. Encrenaz pioneered the study of the molecular cloud in ρ Ophiuchus (Encrenaz, 1974) at the MWO. Somewhat later Michel Guelin spent a postdoctoral at GISS. They encou-raged colleagues in France to apply for time, and a number of them, most prominently, Alain Omont, came with their students. Alain Castets spent a sabbatical with the Texas group in 1984, visiting from the University of Grenoble.

Glenn White from Queen Mary College, London, brought his group on two occasions, once to observe CO with a MWO receiver, and again with a cooled InSb bolometer receiver for submillimeter wavelength observations. Aa Sandqvist came from Sweden and Frank Israel and Thijs de Graauw from the Netherlands. Luis Rodriguez and Jorge Canto were visitors from Mexico.

## 8 CONCLUDING REMARKS

By the 1980s the original partnership had faded away. The Bell Labs group had built its own radio telescope. The Harvard observers were pursuing other interests, among them use of the nearby 14-m millimeter radio telescope of the Five Colleges Radio Astronomy Observatory. And the GISS group had turned to surveying Galactic CO with a 'mini-telescope' of aperture 1m on the roof of the physics building at Columbia University. Larger, more sensitive millimeter wavelength telescopes, built on better sites, were making the MWO less competitive. The discussion of the future of the MWO focused on using it at higher frequencies. The very stable surface combined with the error-correcting subreflector would support such observations. But Mt. Locke lacked the atmospheric transparency for submillimeter observations. In 1985 the University of Texas administration was informed of a plan to move the MWO to Mt. Graham in Arizona, where the University of Arizona was locating telescopes, including a submillimeter telescope of its own. The University of Texas sought private funding to support the move, but was unsuccessful. The next plan was to move the MWO to Mauna Kea, next to the James Clerk Maxwell Telescope, also as a step to submillimeter interferometry. Initial inquiries led to an alternative plan —an agreement with Caltech that the University of Texas become a partner in the Caltech Submillimeter Observatory, which was already sited next to the James Clerk Maxwell Telescope. The MWO would not be moved. The Keck Foundation supported the University of Texas' capital contribution to the Caltech Submillimeter Observatory, and NSF grants supported the research there by the University of Texas. The University of Texas contributions to the operation of the CSO came from the State of Texas through University and the McDonald Observatory. In 1988 the MWO was closed. Later, the University of Texas gave the antenna and astrodome to a research group from the University of Mexico for studies of solar activity and galactic masers. The antenna is to be installed on Sierra Negra, elevation 4600m, a dormant volcano 100 km east of the city of Puebla. Sierra Negra is also the site of the new Large Millimeter Telescope (LMT), a partnership between the University of Massachusetts and the Instituto Nacional de Astrofísica, Óptica y Electrónica in Mexico. With a 50-m aperture, the LMT is the largest single-dish millimeter telescope on Earth. The site is a fitting place for one of the oldest millimeter telescopes to close out its service to the scientific community.

## 9 NOTES

1. Extragalactic CO clouds required larger aperture telescopes; Françoise Combes failed to detect CO in an external galaxy at the MWO and then succeeded using the NRAO 36-ft Radio Telescope.
2. Maryvonne Gerin did not observe in person. Thesis data from the MWO were taken by P. Encrenaz and F. Combes.

## 10 ACKNOWLEDGEMENTS

The authors wish to thank numerous colleagues for assistance in checking facts presented here. Any remaining errors are those of the authors alone.

## 11 REFERENCES

Bash, F.N., and Peters, W.L., 1976. Dynamics of CO molecular clouds in the Galaxy. *Astrophysical Journal*, 205, 786-797.

Bash, F.N., Green, E., and Peters, W.L. III, 1977. The galactic density wave, molecular clouds, and star formation. *Astrophysical Journal*, 217, 464-472.

Beckman, J.E., Watt, C.D., White, G.J., Phillips, J.P., Frost, R. L., and Davis, J.H., 1982. Detection of the 2(1,1) − 2(1,2) rotational emission line of HDO in the Orion Molecular Cloud. *Monthly Notices of the Royal Astronomical Society*, 201, 357-364.

Bennett, J.C., Anderson, J.P., McInnes, P.A., and Whitaker, A.J.T., 1976. Microwave holographic metrology of large aperture antennas. *IEEE Transactions on Antennas and Propagation*, AP-24, 295-303.

Blair, G.N., Evans, N.J., II, Vanden Bout, P.A., and Peters, W.L., III, 1978. The energetics of molecular clouds. II. The S140 molecular cloud. *Astrophysical Journal*, 219, 896-913.

Blitz, L., 1979. The rotation curve of the Galaxy to R = 16 kiloparsecs. *Astrophysical Journal (Letters)*, 231, L115-L119.

Blitz, L., Fich, M., and Stark, A.A., 1982. Catalog of CO radial velocities toward galactic HII regions. *Astrophysical Journal Supplement*, 49, 183-206.

Blitz, L., Magnani, L., and Mundy, L., 1984. High-latitude molecular clouds. *Astrophysical Journal (Letters)*, 282, L9-L12.

Clardy, D.E., and Straiton, A.W., 1968. Radiometric measurements of the Moon at 8.6- and 3.2-millimeter wavelengths. *Astrophysical Journal*, 154, 775-782.

Clegg, R.E.S., and Wootten, H.A., 1980. Circumstellar chlorine chemistry and a search for AlCl. *Astrophysical Journal*, 240, 828-833.

Cogdell, J.R., McCue, J.J.G., Kalachev, P.D., Salomonovich, A.E., Moiseev, I.G., Stacey, J.M., Epstein, E.E., Altschuler, E.E., Geix, G., Day, F.W.B., Hvatum, H., Welch, W.J., and Barath, F.T., 1970. High resolution millimeter reflector antennas. *IEEE Transactions on Antennas and Propagation*, AP-18, 517-527.

Cogdell, J.R., and Davis, J.H., 1973a. Astigmatism in reflector antennas. *IEEE Transactions on Antennas and Propagation*, AP-21, 565-567.

Cogdell, J.R., and Davis, J.H., 1973b. On separating aberrative effects from random scattering effects in radiotelescopes. *Proceedings of the IEEE*, 61, 1344-1345.

Combes, F., Boulanger, F., Encrenaz, P.J., Gerin, M., Bogey, M., Demuynck, C. and Destombes, J.L., 1985. Detection of interstellar CCD. *Astronomy & Astrophysics (Letters)*, 147, L25-L26.

Cudworth, K.M., and Herbig, G., 1979. Two large-proper-motion Herbig-Haro objects. *Astronomical Journal*, 84, 548-551.

Davis, J.H., and Vanden Bout, P., 1973. Intensity calibra-

tion of the interstellar carbon monoxide line at λ2.6 mm. *Astrophysical Letters*, 15, 43-47.

Davis, J.H., Blair, G.N., Van Till, H., and Thaddeus, P., 1974. Vibrationally excited silicon monoxide in the Orion Nebula. *Astrophysical Journal (Letters)*, 190, L117-L119.

Dickman, R.L., 1978. The ratio of carbon monoxide to molecular hydrogen in interstellar dark clouds. *Astrophysical Journal Supplement*, 37, 407-427.

Elmegreen, B.G., and Lada, C.J., 1976. Discovery of an extended (85 pc) molecular cloud associated with the M17 star-forming complex. *Astronomical Journal*, 81, 1089-1094.

Elmegreen, D.M., and Elmegreen, B.G., 1979. CO and near IR observations of a filamentary cloud, L43. *Astronomical Journal*, 84, 615-620.

Encrenaz, P., 1974. A new source of intense molecular emission in the Rho Ophiuchi complex. *Astrophysical Journal (Letters)*, 189, L135-L136.

Erickson, N.R., 1977. A directional filter diplexer using optical techniques for millimeter to submillimeter wavelengths. *IEEE Transactions on Millimeter Theory and Techniques*, MTT-32, 865-866.

Evans, N.J., II, Blair, G.N., and Beckwith, S., 1977. The energetics of molecular clouds. I. Methods of analysis and application to the S 255 molecular cloud. *Astrophysical Journal*, 217, 449-463.

Evans, N.J., II, Plambeck, R.L., and Davis, J.H., 1979. Detection of the $3_{12}-2_{11}$ transition of interstellar formaldehyde at 1.3 millimeters. *Astrophysical Journal (Letters)*, 227, L25-L28.

Evans N.J., II, and Blair, G.N., 1981. The energetic of molecular clouds. III. The S235 molecular cloud. *Astrophysical Journal*, 246, 394-408.

Evans, N.J., II, Blair, G.N., Harvey, P., Israel, F., Peter, W.L., III, Scholtes, M., de Graauw, T., and Vanden Bout, P., 1981. The enegetics of molecular clouds. IV. The S88 molecular cloud. *Astrophysical Journal*, 250, 200-212.

Evans, N.J., II, Blair, G.N., Nadeau, D., and Vanden Bout, P., 1982. The energetics of molecular clouds. V. The S37 molecular cloud. *Astrophysical Journal*, 253, 115-130.

Goldsmith, P.F., 1988. *Instrumentation and Techniques for Radio Astronomy*. New York, IEEE Press.

Gottlieb, C.A., and Ball, J.A., 1973. Interstellar sulfur monoxide. *Astrophysical Journal (Letters)*, 184, L59-L64.

Guelin, M., Langer, W., Snell, R., and Wootten, H.A., 1977. Observations of DCO⁺: the electron abundance in dark clouds. *Astrophysical Journal (Letters)*, 217, L165-L168.

Irvine, W.M., Abraham, Z., A'Hearn, M., Altenhoff, W., Andersson, Ch., et al., 1984. Radioastronomical observations of comets IRAS-Araki-Alcock (1983d) and Sugano- Saigusa-Fujikawa (1983e). *Icarus*, 60, 215-220.

Kutner, M.L., Tucker, K.D., Chin, G. and Thaddeus, P., 1977. The molecular complexes in Orion. *Astrophysical Journal*, 215, 521-528.

Lada, C.J., 1976. Detailed observations of the M17 molecular cloud complex. *Astrophysical Journal Supplement Series*, 32, 603-629.

Lada, C.J., Elmegreen, B.G., Cong, H-I, and Thaddeus, P., 1978. Molecular clouds in the vicinity of W3, W4, and W5. *Astrophysical Journal (Letters)*, 226, L39-L42.

Lada, C.J., 1985. Cold outflows, energetic winds, and enigmatic jets around young stellar objects. *Annual Reviews of Astronomy and Astrophysics*, 23, 267-317.

Lambert, D.L., and Vanden Bout, P.A., 1978. Constraints on the properties of circumstellar shells from observations of thermal CO and SiO millimeter line emission. *Astrophysical Journal*, 221, 854-860.

Liszt, H.S., and Vanden Bout, P.A., 1985. Upper limits on the O2/CO ratio in two dense interstellar clouds. *Astrophysical Journal*, 291, 178-182.

Loren, R.B., Vanden Bout, P.A., and Davis, J.H., 1973. Car-

Carbon monoxide emission from nebulosity associated with Herbig Be and Ae type stars. *Astrophysical Journal (Letters)*, 185, L67-L70.

Loren, R.B., Peters, W.L., and Vanden Bout, P.A., 1974. Collapsing molecular clouds? *Astrophysical Journal (Letters)*, 194, L103-L107.

Loren, R.B., 1977. The Monoceros R2 cloud: near infrared and molecular observations of a rotating collapsing cloud. *Astrophysical Journal*, 215, 129-150.

Loren, R.B. and Mundy, L.G., 1984. The methyl cyanide hot and warm cores in Orion: statistical equilibrium excitation models of a symmetric top molecule. *Astrophysical Journal*, 286, 232-251.

Loren, R.B., and Wootten, H.A., 1986. Submillimeter molecular spectroscopy with the Texas Millimeter Wave Observatory Radio Telescope. *Astrophysical Journal*, 306, 889-906.

Loren, R.B., Wootten, H.A., and Wilking, B.A., 1990. Cold DCO⁺ cores and protostars in the warm Rho Ophiuchi cloud. *Astrophysical Journal*, 365, 269-286.

Magnani, L., Blitz, L., and Mundy, L., 1985. Molecular gas at high galactic latitudes. *Astrophysical Journal*, 295, 402-421.

Mangum, J. G., Wootten, H.A., Loren, R. B., and Wadiak, E.J., 1990. Observations of the formaldehyde emission in Orion-KL - abundances, distribution, and kinematics of the dense gas in the Orion molecular ridge. *Astrophysical Journal*, 348, 542-556.

Mayer, C.E., Davis, J.H., Peters, W.L.III, and Vogel, W., 1983. A holographic surface measurement of the Texas 4.9-m antenna at 86 GHz. *IEEE Transactions on Instrumentation and Measurement*, IM-32, 102-109.

Mundy, L.G., Snell, R.L., Evans, N.J., II, Goldsmith, P.F., and Bally, J., 1986. Models of molecular cloud cores. II - Multitransition study of C³⁴S. *Astrophysical Journal*, 306, 670-681.

Mundy, L.G., Evans, N.J., II, Snell, R.L., and Goldsmith, P.F., 1987. Models of molecular cloud cores. III. A multitransition study of $H_2CO$. *Astrophysical Journal*, 318, 192-409.

Penzias, A.A., and Burrus, C.A., 1973. Millimeter-wavelength radio-astronomy techniques. *Annual Review of Astronomy and Astrophysics*, 11, 51-72.

Sahai, R., Wootten, A., and Clegg, R.E.S., 1984. SiS in circumstellar shells. *Astrophysical Journal*, 284, 144-156.

Sandqvist, Aa., 1989. 2-mm $H_2CO$ emission in the Sgr A molecular complex at the Galactic Center. *Astronomy & Astrophysics*, 223, 293-303.

Scott, P.F., and Ryle, M., 1977. A rapid method of measuring the figure of a radio telescope reflector. *Monthly Notice of the Royal Astronomical Society*, 178, 539-544.

Snell, R.L. and Wootten, H.A., 1977. Detection of interstellar DNC. *Astrophysical Journal (Letters)*, 216, L111-L114.

Snell, R.L. and Wootten, H.A., 1979. Observations of interstellar HNC, DNC, and HN¹³C: temperature effects on deuterium fractionation. *Astrophysical Journal (Letters)*, 228, L748-L754.

Snell, R.L., Loren, R.B., and Plambeck, R.L., 1980. Observations of CO in L1551: evidence for stellar wind driven shocks. *Astrophysical Journal (Letters)*, 239, L17-L22.

Snell, R.L., Langer, W.D., and Frerking, M.A., 1982. Determination of density structure in dark clouds from CS observations. *Astrophysical Journal*, 255, 149-159.

Snell, R.L., Mundy, L.G., Goldsmith, P.F., Evans, N.J., II, and Erickson, N.R., 1984. Models of molecular clouds. I. Multitransition study of CS. *Astrophysical Journal*, 276, 225-245.

Snyder, L.E., and Buhl, D., 1974. Detection of possible maser emission near 3.48 millimeters from an unidentified molecular species in Orion. *Astrophysical Journal (Letters)*, 189, L31-L33.

Strom, S.E., Grasdelen, G.L., and Strom, K.L., 1974. Infra-

red and optical observations of Herbig-Haro objects. *Astrophysical Journal*, 191, 111-142.

Thaddeus, P., Mather, J., Davis, J.H., and Blair, G.N., 1974. Detection of the J=1-0 rotational transition of vibrationally excited silicon monoxide. *Astrophysical Journal (Letters)*, 192, L33-L36.

Tolbert, C.W., and Straiton, A.W., 1964. 35-Gc/s, 70-Gc/s, and 94 Gc/s Cytherean radiation. *Nature*, 204, 1242-1245.

Tolbert, C.W., and Straiton, A.W., 1965. Investigation of Tau A and Sgr A millimeter wavelength radiation. *Astronomical Journal*, 70, 177-180.

Tolbert, C.W., 1965. Millimetre wave-length spectra of the Crab and Orion Nebulae. *Nature*, 206, 1304-1307.

Tolbert, C.W., Straiton, A.W., and Krause, L.S., 1965. A 16-foot diameter millimeter wavelength antenna system. Its characteristics and its applications. *IEEE Transactions on Antennas and Propagation*, AP-13, 225-229.

Tolbert, C.W., 1966. Observed millimeter wavelength brightness temperatures of Mars, Jupiter, and Saturn. *Astronomical Journal*, 71, 30-32.

Tucker, K.S., Kutner, M.L., and Thaddeus, P., 1973. A large carbon monoxide cloud in Orion. *Astrophysical Journal (Letters)*, 186, L13-L17.

Ulich, B.L., Cogdell, J.R., and Davis, J.H., 1973. Planetary brightness temperatures at 8.6 mm and 3.1 mm wavelengths. *Icarus*, 19, 59-82.

Ulich, B.L., 1974. Absolute brightness temperature measurements at 2.1 mm wavelength. *Icarus*, 21, 254-261.

Ulich, B.L., Davis, J.H, Rhodes, P.J., and Hollis, J.M., 1980. Absolute brightness temperature measurements at 3.5-mm wavelength. *IEEE Transactions on Antennas and Propagation*, AP-28, 367-377.

Vanden Bout, P.A., Mundy, L.G., Davis, J.H., Loren, R.B., and Butner, H., 1985. Calibration of millimeter-wavelength spectral lines – effect of harmonic mixer response. *Astrophysical Journal*, 295, 139-142.

Van Den Eeden, S.K., Tanner, C.M., Bernstein, A.L., Fross, R.D., Leimpeter, A., Bloch, D.A., and Nelson, L.M., 2003. Incidence of Parkinson's Disease: variations by age, gender, and race/ethnicity. *American Journal of Epidemiology*, 157, 1015-1022.

Wilson, R.W., Jefferts, K.B., and Penzias, A.A., 1970. Carbon monoxide in the Orion Nebula. *Astrophysical Journal (Letters)*, 161, L43-L44.

Wootten, H.A., 1977. The molecular cloud associated with the supernova remnant W44. *Astrophysical Journal*, 216, 440-445.

Wootten, H.A., Snell, R.L., and Glassgold, A.E., 1979. The determination of electron abundances in interstellar clouds. *Astrophysical Journal*, 234, 876-880.

Wootten, A., 1981. A dense molecular cloud impacted by the W28 supernova remnant. *Astrophysical Journal*, 245, 105-114.

Wootten, H.A., Loren, R.B., and Snell, R.L., 1982. A study of DCO$^+$ emission regions in interstellar clouds. *Astrophysical Journal*, 255, 160-175.

Ziurys, L.M., Saykally, R.J., Plambeck, R.L., and Erickson, N.R., 1982. Detection of the N = 3-2 transition of CCH in Orion and determination of the molecular rotational constants. *Astrophysical Journal*, 254, 94-99.

Paul Vanden Bout is an Emeritus Scientist and former Director of the U.S. National Radio Astronomy Observatory. His research interests are centered on interstellar matter: molecular spectroscopy at millimeter wavelengths, star formation, and molecular line emission at high redshifts. While Director he supported the establishment of the NRAO Archives and in retirement he is devoting some of his time to the history of radio astronomy.

John Davis is a Professor of Electrical Engineering (retired) at the University of Texas in Austin. His research interests include the evaluation of antenna surfaces and microwave instrumentation, particularly as applied to millimeter wavelength radio astronomy.

Robert Loren is a former Research Astronomer at the University of Texas in Austin. He is the author of numerous articles based on molecular spectroscopy of the interstellar medium, with a prominent theme being the connection between carbon monoxide emission and star formation. He is a co-discoverer of the first molecular outflow source.